\documentclass[%
 reprint,
 amsmath,amssymb,
 aps,
prb,
floatfix,
]{revtex4-2}
\usepackage{graphicx}
\usepackage{dcolumn}
\usepackage{bm}
\usepackage{xcolor}

\begin{document}

\preprint{APS/123-QED}

\title{Inelastic neutron scattering investigation of the crystal field excitations of NdCo$_5$}

\author{F. de Almeida Passos$^{1}$}
\author{G. J. Nilsen$^{2,3}$}
\author{C. E. Patrick$^4$}
\author{M. D. Le$^2$}
\author{G. Balakrishnan$^5$}
\author{Santosh Kumar$^{5,6}$}
\altaffiliation[Present address: ]{Department of Physics,
Indian Institute of Technology Dharwad, Karnataka 580011, India}
\author{A. Thamizhavel$^6$}
\author{D. R. Cornejo$^7$}
\author{J. Larrea Jim\'{e}nez$^{1}$}
\email{corresponding author: larrea@if.usp.br}

 \affiliation{$^1$Laboratory for Quantum Matter under Extreme Conditions, Institute of Physics, University of São Paulo, São Paulo, Brazil}
 \affiliation{$^2$ISIS Pulsed Neutron and Muon Source, STFC Rutherford Appleton Laboratory, Didcot OX11 0QX, UK}
 \affiliation{$^3$Department of Mathematics and Physics, University of Stavanger, 4036 Stavanger, Norway}
 \affiliation{$^4$Department of Materials, University of Oxford, Oxford OX1 3PH, United Kingdom}
 \affiliation{$^5$Department of Physics, University of Warwick, Coventry CV4 7AL, United Kingdom}
 \affiliation{$^6$Department of Condensed Matter Physics and Materials Science, Tata Institute of Fundamental Research, Mumbai 400005, India}
 \affiliation{$^7$Institute of Physics, University of São Paulo, São Paulo, Brazil}

\date{\today}

\begin{abstract}
We present an inelastic neutron scattering study of the crystal electric field (CEF) levels in the intermetallic ferrimagnets  RECo$_{5}$ (RE = Nd and Y). In NdCo$_{5}$, measurements at $5~$K reveal two levels at approximately 28.9 and 52.9 meV. Crystal field calculations including the exchange field $B_{\textrm{exc}}$ from the Co sites account for both of these, as well as the spectrum at temperatures above the spin-reorientation transition at $\sim 280$~K. In particular, it is found that both a large hexagonal crystal field parameter $A_{6}^6\langle r^6 \rangle$ and $B_{\textrm{exc}}$ are required to reproduce the data, with the latter having a much larger value than that deduced from
previous computational and experimental studies. Our study sheds light on the delicate interplay of terms in the rare-earth Hamiltonian of RECo$_5$ systems, and is therefore expected to stimulate further experimental and computational work on the broader family of rare-earth permanent magnets.

\end{abstract}


\maketitle


\section{Introduction}
The rare-earth (RE) intermetallics RECo$_5$ have been extensively studied in the last three decades due to their attractive magnetic properties, which include high saturation magnetization and ordering temperature $T_C$, as well as strong magnetic anisotropy and large coercivity \cite{strnat}. These can be understood as arising from the features of the rare-earth and transition-metal (TM) sub-lattices: the large saturation magnetization and high $T_C$ are generated by the strongly interacting itinerant \textit{d}-electrons on the TM sub-lattice, while the localized RE \textit{f}-electrons, crystal electric field (CEF), and exchange field ($B_{\textrm{exc}}$) from the TM site together produce the magnetic anisotropy \cite{richter}. 

Despite the fact that the CEF plays an important role in the mechanisms that underlie the magnetic properties of RECo$_5$ systems, the accurate determination of crystal field parameters (CFPs) in RECo$_5$ remains a challenge. Theoretical calculations based on \textit{ab-initio} methods have produced a wide range of CFPs and exchange fields \cite{pourovskii,patrick}, with a much narrower range of predicted physical properties, rendering comparisons with experiment ambiguous. Inelastic neutron scattering (INS) is one of the best tools to obtain both sets of parameters, but has only so far been applied to a few members of the RECo$_5$ family. This is at least in part because the exchange field $B_{\textrm{exc}}$ both fully splits the RE ground state multiplet and mixes in higher multiplets, resulting in highly complex spectra. The availability of inelastic neutron scattering data is nevertheless expected to help to distinguish between theoretical parameter sets, and thus to identify the most promising theoretical tools to design the next generation of permanent magnets.


With this aim in mind, we here focus on the CEF in the NdCo$_5$ compound, which crystallizes in the hexagonal ($P6/mmm$) space group symmetry with lattice parameters $a=$ 5.0200(9) \AA, and $c=$ 3.9664(4) \AA, respectively \cite{wang}. Several studies using magnetization and neutron diffraction have already been performed on this compound: in particular, a spin-reorientation transition (SRT) between $T_{SR1}=240$ K and $T_{SR2}=280$~K \cite{klein} and a magnetic moment smaller than the expected saturation value have been observed \cite{alameda}. Regarding the CEF, parameters from a range of theoretical calculations have been found to be broadly compatible with magnetization and other bulk data \cite{pourovskii,patrick}, with the best agreement at low temperature being obtained using dynamical mean field theory (DMFT) \cite{pourovskii}. The latter work suggests that a strong hybridization between the Nd $4f$ and Co $3d$ orbitals generates a large $6^{\textnormal{th}}$ order CEF coefficient $A_{6}^6\langle r^6\rangle$, which in turn increases the easy plane anisotropy and reduces the low-temperature ordered moment on the Nd site.

Our study completes the picture of the CEF in NdCo$_5$ via inelastic neutron scattering experiments on both it and the isostructural compound YCo$_5$, where the excitation spectrum is dominated by phonons. In NdCo$_5$, two excitations at $28.9$~meV and $52.9$~meV are clearly observed at 5 K. Using previous calculations as a starting point, we fit the full inelastic neutron scattering spectrum to extract a set of CFPs and $B_{\textrm{exc}}$ that explain the observed CEF excitations, including the spectrum above the spin-reorientation transition at 300 K. Remarkably, we find a much larger $B_{\textrm{exc}}$ than previous \textit{ab-initio} calculations, as well as an $A_{6}^6\langle r^6\rangle$ coefficient in good agreement with the DMFT calculations discussed above.

\section{Methods}
\subsection{Experimental}
Polycrystalline ingots of NdCo5 and YCo5 were synthesized by arc melting high purity Nd, Y and Co elements in stoichiometric proportions on a water-cooled copper hearth in an argon atmosphere. The as-cast ingots were then ground to powder form for the neutron experiments. The phase purity of the powders was checked using powder x-ray diffraction, prior to the neutron measurements. 

The inelastic neutron scattering measurements for both samples were performed on the MARI spectrometer at the ISIS Neutron and Muon Source, UK. Two sets of incident neutron energies were selected using a Fermi chopper and repetition rate multiplication: $E_i=180/30$~meV, and $E_i=80/11$~meV. The corresponding resolutions at the elastic line were $7/0.7$ meV and $3.8/0.3$~meV, respectively. Data was collected at $5$~K and $300$~K in both configurations, and corrected for $k_i/k_f$ to yield the dynamical structure factor $S(|Q|,\Delta E = \hbar\omega)$.

\subsection{Crystal Field Hamiltonian}
The Hamiltonian used to fit the NdCo$_5$ spectra was:
\begin{align}
    \mathcal{H}=&\lambda\mathbf{L}\cdot\mathbf{S}+2\mu_B\mathbf{B}_{\textrm{exc}}\cdot\mathbf{S}+\mathcal{H}_{\textrm{cf}}
    \label{eq:hamiltonian}
\end{align}
\noindent where the first term represents the spin-orbit coupling, the second the coupling between the exchange field and localized rare earth spin moment, the third the Zeeman energy, and the fourth the crystal field Hamiltonian. By choosing the quantization axis along the hexagonal $c$ axis, the exchange field is taken to be parallel to the $x$ axis (crystallographic $a$-axis) below the spin-reorientation transition, and parallel to the $z$ axis (crystallographic $c$-axis) above it. For $f$ electrons and the $6/mmm$ site symmetry of the Nd atoms, four crystal field parameters are allowed: following the notation used in the RECo$_5$ literature, these are denoted $A_{2}^0\langle r^2 \rangle$, $A_{4}^0\langle r^4 \rangle$, $A_{6}^0\langle r^6 \rangle$, and $A_{6}^6\langle r^6 \rangle$. The Stevens and Wybourne conventions are related 
via $A_{k}^q\langle r^k \rangle = \lambda_{kq}W_{k}^q$, where $\lambda_{kq}$ are multiplicative tabulated factors \cite{spectre}. The crystal field Hamiltonian then reads: 
\begin{align}
    \mathcal{H}_{\textrm{cf}} = & \Theta_2 A_{2}^0\langle r^2 \rangle \hat{O}_2^0 + \Theta_4 A_{4}^0\langle r^4 \rangle \hat{O}_4^0 \nonumber \\ 
    &+ \Theta_6 \left[ A_{6}^0\langle r^6 \rangle \hat{O}_6^0 + A_{6}^6\langle r^6 \rangle \hat{O}_6^6 \right]
\end{align}
\noindent where $\hat{O}_{k}^q$ are the Stevens operator equivalents. The dynamical structure factor $S(|Q|,\Delta E)$ was evaluated using the standard expression for a powder in the dipole approximation:
\begin{align}
    S(|Q|&,\Delta E) = \frac{2}{3}\left(\frac{\gamma r_0}{2}\right)^2g^2f(|Q|)^2\sum_{\nu}p_\nu \nonumber \\ 
    &\times \sum_{\nu^\prime}\sum_{\alpha=\{x,y,z\}}\left|\langle \nu^\prime |J_\alpha| \nu \rangle\right|^2 \delta(E_{\nu^\prime}-E_{\nu}- \Delta E),
\end{align}
 \noindent where $p_{\nu}$ is the Boltzmann population factor for initial state $|\nu\rangle$ in the $|SLJm_J\rangle$ basis, and $f(|Q|)^2$ is the Nd form factor. Given the large dimension of the parameter space, the least-squares fits were initialized using three literature parameter sets \cite{zhao,pourovskii,patrick} (Table \ref{tab:CFPs}) as well as the parameters obtained by performing a grid search in the lower dimensional space $\{|B_{\textrm{exc}}|,A_{2}^0\langle r^2 \rangle,A_{6}^6\langle r^6 \rangle\}$ of the parameters that all previous calculations identify as most significant.

\section{Results and Discussion}

\subsection{Background subtraction}

\begin{figure*}
\includegraphics[width=\textwidth]{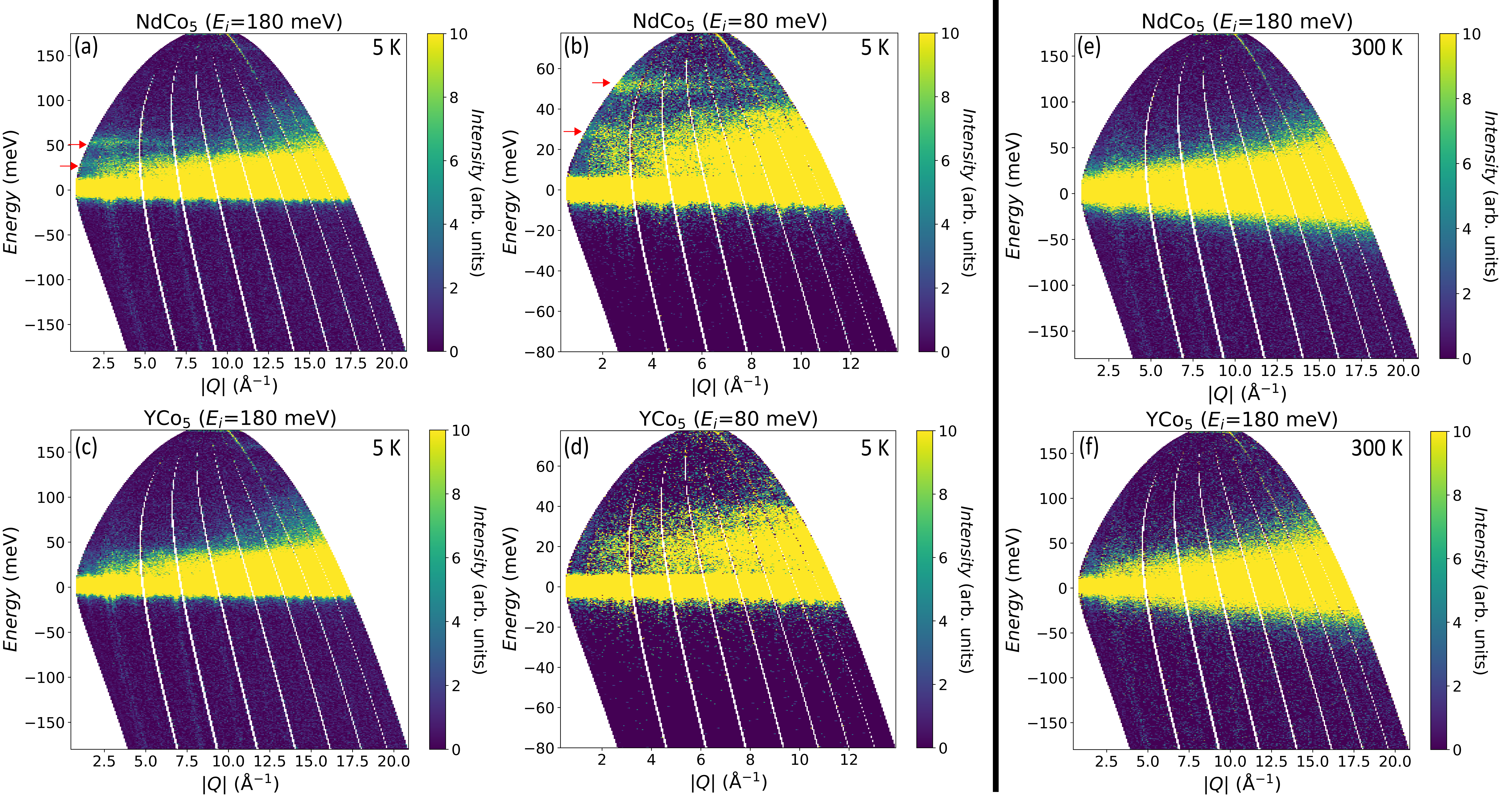}
\caption{\label{fig:colorplots} Experimental inelastic neutron scattering intensity spectra of: (a) NdCo$_5$ at 5 K obtained using an incident neutron energy of E$_i$=180 meV; (b) NdCo$_5$ at 5 K obtained with E$_i$=80 meV; (c) YCo$_5$ at 5 K obtained with E$_i$=180 meV; (d) YCo$_5$ at 5 K obtained with E$_i$=80 meV; (e) NdCo$_5$ at 300 K obtained with E$_i$=180 meV, and (f) YCo$_5$ at 300 K obtained with E$_i$=180 meV. The red arrows in (a) and (b) indicate the two observed CEF excitations on the NdCo$_5$ spectra.}
\label{fig:allspec}
\end{figure*}

The dynamical structure factors $S(|Q|,\Delta E)$ of NdCo$_5$ and YCo$_5$ at $5$~K and $E_i=180$ and $80$~meV are shown in Figs. \ref{fig:colorplots}(a-d). At large wavevector transfers $|Q|$, the spectra of both materials are dominated by phonons, which appear in several strong bands between $15$ and $60$~meV. In the case of YCo$_5$, no other features are observed in the $(|Q|,\Delta E)$ range of our experiments: the magnons expected from the magnetic order are either too weak or too broad to observe. On the other hand, the $E_i = 80$~meV and $180$~meV spectra of NdCo$_5$ (Figs. \ref{fig:colorplots}(a) and (b)) reveal two features at $\sim50$~meV and $\sim 30$~meV (see arrows), henceforth denoted as the high energy (HE) and low energy (LE) features. Both have $|Q|$-dependences that are apparently consistent with magnetic excitations. Turning to the $300$~K data (Figs. \ref{fig:colorplots}(e) and (f)), $S(|Q|,\Delta E)$ for YCo$_5$ continues to be dominated by phonon scattering, while both of the lines observed for NdCo$_5$ at low temperature are absent from the spectrum. This drastic change will be shown to result from the spin-reorientation transition that switches the magnetization easy axis from the $a$ axis to the $c$ axis at $T_{SR1}$ and $T_{SR2}$.

\begin{figure*}
\includegraphics[width=0.4\textwidth]{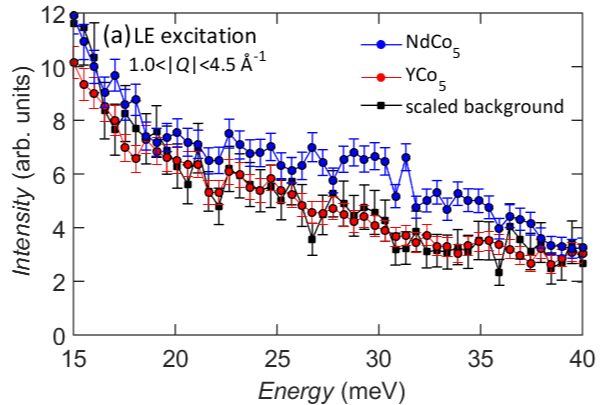}
\includegraphics[width=0.4\textwidth]{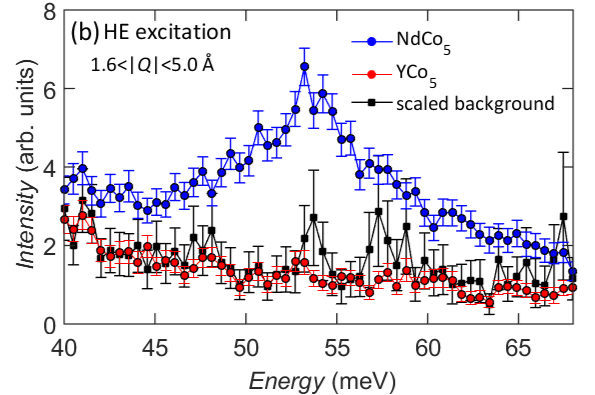}
\includegraphics[width=0.4\textwidth]{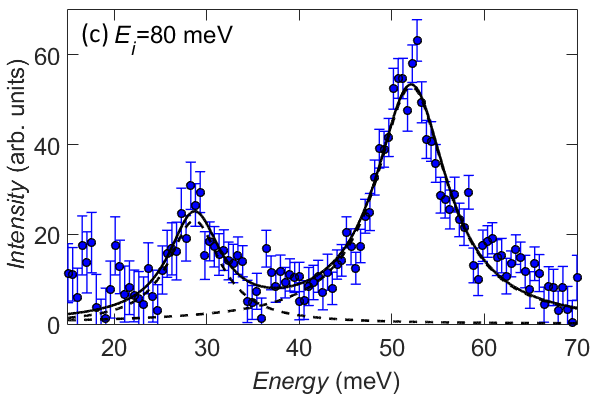}
\includegraphics[width=0.4\textwidth]{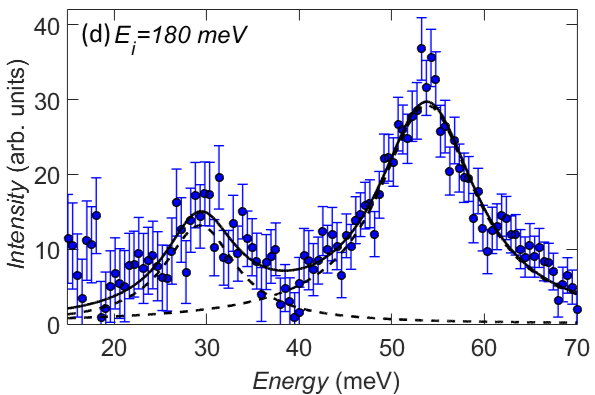}
\caption{\label{fig:background} (a) Dynamical structure factor $S(|Q|,\Delta E)$ of NdCo$_5$ and YCo$_5$ at 5 K obtained by integrating the $E_i=180$~meV data over a Q range of 1.0$<$Q$<$4.5 \r{A}$^{-1}$ showing the peak in the NdCo$_5$ spectrum corresponding to the LE CEF excitation. (b) Energy spectra of NdCo$_5$ and YCo$_5$ at 5 K obtained integrating the $E_i=180$~meV data over a Q range of 1.6$<$Q$<$5.0 \r{A}$^{-1}$ showing the peak in the NdCo$_5$ spectrum corresponding to the HE CEF excitation. (c) and (d) NdCo$_5$ background-subtracted spectrum using $E_i=80$ and $180$~meV, respectively, after intensity correction to account for the different $|Q|$ integration ranges (filled circles) fitted by a Lorentzian function (solid line).}
\label{fig:ecuts}
\end{figure*}

Before analyzing the spectra in detail, the CEF component of the scattering must first be isolated from the remainder. To achieve this, we compare two different approaches: subtracting either the YCo$_5$ data from the NdCo$_5$ data or using the scaled high-$|Q|$ phonon spectrum. Firstly, despite the difference in mass between Y and Nd, YCo$_5$ has a very similar phonon spectrum to NdCo$_5$ (see Fig. \ref{fig:allspec}), with only a slight shift in phonon frequencies at low energies. The relative intensities are furthermore nearly identical across the whole energy range due to the scattering lengths of Y and Nd being very close in magnitude ($b_\textrm{Y} = 7.75$fm and $b_{\textrm{Nd}}=7.69$fm). In addition, at sufficiently high-$|Q|$ the magnetic contribution should be negligible compared to the phonon scattering. Therefore, we also evaluate the phonon contribution to the NdCo$_5$ spectrum assuming that its low-$|Q|$ and high-$|Q|$ phonon scattering ratio scales in the same manner as in the isostructural compound YCo$_5$ \cite{murani}, where the RE site is non-magnetic. Figs. \ref{fig:ecuts}(a) and (b) show excellent agreement between the NdCo$_5$ phonon background calculated using this scale function and the YCo$_5$ spectrum, making both suitable for removing the non-magnetic contribution. The subtracted spectra, shown along $\Delta E$ in Figure \ref{fig:ecuts}(c) and (d) and $|Q|$ in Figure \ref{fig:form} indicate that the phonon contribution is cleanly removed at energies above $20$~meV. 

In order to verify the magnetic origin of the observed features, we begin by analyzing the $|Q|$-dependence of the two CEF excitations in NdCo$_5$. Integrating the background-subtracted $S(|Q|,\Delta E)$ over the energy transfer ranges of both, we obtain their $|Q|$-dependence, as shown in Fig. \ref{fig:form}. The red line in Fig. \ref{fig:form} is the squared magnetic form factor $f(|Q|)^2$ for the Nd$^{3+}$ ion calculated in the dipole approximation $f(|Q|)=\langle j_0 \rangle+c_2\langle j_2 \rangle$ with $c_2 = (2-g_J)/g_J$ \cite{formfactor}, where $g_J$ in the Landé g-factor. We can see that the intensities for both NdCo$_5$ CEF excitations decrease with $|Q|$, as expected for magnetic scattering, and that they also agree well with the Nd$^{3+}$ squared form factor, even at high $|Q|$, where the phonon intensity dominates. This provides additional reassurance that the background subtraction cleanly isolates the CEF magnetic scattering contribution, as well as showing that the strong $f-d$ hybridization suggested in Ref. [\onlinecite{pourovskii}] is nearly isotropic.

{\subsection{Extracting CFPs}

We now turn to cuts of $S(|Q|,\Delta E)$ along $\Delta E$ in the energy ranges $20<\Delta E<40$~meV and $40<\Delta E<70$~meV in Fig. \ref{fig:background}(a) and (b): these were obtained by integrating the $E_i = 180$~meV data at 5 K over the $|Q|$-ranges $1.0<Q<4.5$~\AA{}$^{-1}$, $1.6<Q<5.0$~\AA{}$^{-1}$, respectively. Since the cuts run over different $|Q|$ ranges, the intensity was corrected by dividing it by the ratio of $f(|Q|)^2$ integrated over the $|Q|$ ranges above and $\int_{0}^{\infty}f(|Q|)^2dQ$. This is justified by the fact that the $|Q|$-dependences in Figs. \ref{fig:form}(a) and (b) are in good agreement with $f(|Q|)^2$ for Nd$^{3+}$. Firstly, it is evident that both the LE and HE features have a roughly Lorentzian profile and are considerably broader than the (Gaussian) instrumental resolution, which is estimated to be $5.8$ meV for the LE peak and $5.2$ meV for the high-energy HE peak for the $E_i=180$~meV data, and $2.5$ and $1.7$ meV for the LE and HE peaks, respectively, for the $E_i=80$~meV data. They are also broad compared to the CEF excitations in SmCo$_5$ \cite{tils}. This justifies the choice of a model containing two Lorentzians to fit the experimental data.

This broadening of the CEF excitations has several possible origins: \textit{(i)} dispersion due to long-range interactions between the localized Nd$^{3+}$ $f$-electrons mediated by the conduction electrons \cite{jensen}; \textit{(ii)} magneto-elastic coupling between the CEF excitations and the phonons \cite{fulde}; and \textit{(iii)} $f-d$ exchange between the localized $f$-electrons and the itinerant $d$ electrons (Landau damping), in a manner analogous to the $f-s$ broadening mechanism proposed in \cite{becker}. Since the current experiments were performed on polycrystalline samples, we were unable to resolve the dispersion of the CEF excitations, although powder averaging would be unlikely to produce the Lorentzian lineshape observed experimentally. This means that \textit{(i)} is almost certainly not the dominant source of broadening. For \textit{(ii)}, the similarity of the phonon spectra of YCo$_5$ and NdCo$_5$ at all $|Q|$ suggests that any magneto-elastic effects present are too small to explain the large broadening of both the LE and HE Features. Finally, regarding \textit{(iii)}, significant broadening effects have been observed in several other itinerant rare earth systems, where they were ascribed to coupling between the localised $4f$ moments and electron-hole excitations in the valence $5s$-band \cite{becker}. If a similar mechanism couples the Nd moments to the Co $3d$-band in NdCo$_5$, we expect that only the temperature-independent term $\propto K_{ex}\mathcal{N}(0)$ is active, as the Kramers degeneracy is broken by the exchange field.

\begin{figure}[h]
    \centering
    \includegraphics[width=0.42\textwidth]{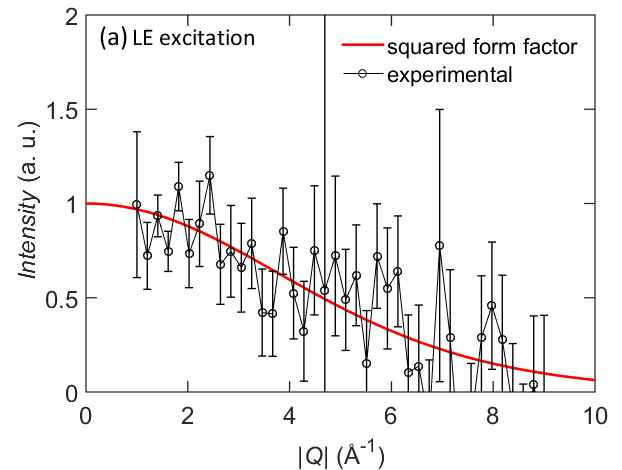}
    \includegraphics[width=0.42\textwidth]{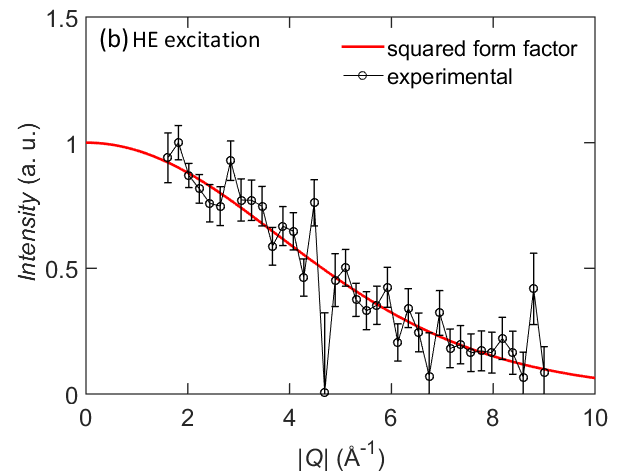}
    \caption{Normalized scattered intensities as a function of $|Q|$ of NdCo$_5$ spectrum measured with $E_i$=180 meV at 5 K obtained integrating over the (a) LE peak range (20 to 40 meV) and (b) HE peak range (40 to 70 meV). The solid red line show the calculated squared form factor for the Nd$^{3+}$ ion using the analytical approximation \cite{formfactor}.}
    \label{fig:form}
\end{figure}

The NdCo$_5$ background-subtracted spectra for both the $E_i=80$ and $180$~meV data were thus fitted to two Lorentzian functions ($L_1$ and $L_2$) reproducing the intensity $I(\Delta E)$:

\begin{align}
    I(\Delta E) = L_1 (\Delta E; \Delta E_1, \gamma_1) + L_2 (\Delta E; \Delta E_2, \gamma_2),
\end{align}

\noindent where the widths of Lorentzians, $\gamma_1$ and $\gamma_2$ were allowed to vary freely for the two peaks at positions $\Delta E_1$ and $\Delta E_2$, respectively. Attempts to fit three Lorentzian were unstable. Table \ref{tab:fit} shows the peak parameters for both spectra.

\begin{table}[ht]
\caption{\label{tab:fit}
Fitted Lorentzian function parameters for each peak in NdCo$_5$ spectrum at 5 K for both neutron incident energies $E_i=80$ and $E_i=180$~meV.
}
\begin{ruledtabular}
\begin{tabular}{lccc}
\textrm{$E_i=80$~meV}&
\textrm{Center (meV)}&
\textrm{Intensity}&
\textrm{FWHM (meV)}\\
\colrule
Low-energy peak & 28.6(4) & 249(28) & 7(1)\\
High-energy peak & 52.1(2) & 785(26) & 9.4(4)\\
\hline
\hline
$E_i=180$~meV &   &   &   \\
\colrule
Low-energy peak & 29.3(5) & 196(23) & 9(1)\\
High-energy peak & 53.8(2) & 590(20) & 12.8(6)\\
\end{tabular}
\end{ruledtabular}
\end{table}

The fits of the model parameters from the Hamiltonian given in Section IIB were carried out using a custom \texttt{python} code, and \texttt{SPECTRE} and \texttt{PyCrystalField} software packages \cite{spectre,scheie}. In the first case, the basis was truncated to the three lowest $J$ multiplets of the $^4I$ term ($J=9/2,11/2,13/2$), giving 36 basis states. For the \texttt{SPECTRE} fits, both the ground state $^4I$ and excited $^4F$ terms were considered. Since \texttt{SPECTRE} uses the Wybourne operator equivalents $\hat{C}_q^k$ and coefficients $W_k^q$, the latter were converted to $A_k^q\langle r^k \rangle$ using tabulated factors. In both cases, the spin-orbit coupling constant $\lambda$ was chosen to be $540$~K \cite{patrick}, and the average peak positions and intensities of the $E_i=80$ and $E_i=180$~meV data for the LE and HE features at $5$~K were used to fit the CEF Hamiltonian parameters. On the other hand, for the \texttt{PyCrystalField} fits, only the $J=9/2$ ground state term was considered. However, \texttt{PyCrystalField} has the advantage of fitting the whole neutron spectrum instead of fitting only the transition energies and relative intensities. During the fitting procedure, the neutron spectrum was found to be strongly sensitive to just three parameters: the $A_{2}^0\langle r^2 \rangle$ and $A_{6}^6\langle r^6 \rangle$ CFPs, and the exchange field $B_\textrm{exc}$. Due to the large width of the peaks -- which renders reliable extraction of intensities challenging -- and the small apparent number of measured levels, the \texttt{PyCrystalField} was found to provide the most stable and consistent fits to the data, and the remaining discussion will center around the parameters extracted from these.

Table \ref{tab:CFPs} shows the values obtained for the CFPs as well as the magnitude of the exchange field $B_{\textrm{exc}}$, whose direction is along the $a$-axis below the lower spin-reorientation transition temperature $T_{SR1}$. Although there are some inconsistencies between the fitted parameters, all three approaches suggest a large $A_{6}^6\langle r^6 \rangle$ and $B_\textrm{exc}$. In the Appendix (Table \ref{tab:eigenvectors}), we also show the eigenvalues with respective eigenvector coefficients in the $|SLJm_J\rangle$ basis obtained from the \texttt{PyCrystalField} fitting with the quantization axis chosen along the $a$ direction for better comparison with previous references \cite{alameda,pourovskii}. Figure \ref{fig:comp_spectrum} depicts the comparison between intensities using CFPs obtained from previous works \cite{patrick, zhao, pourovskii} and from our fitting using \texttt{PyCrystalField}. Of the spectra calculated using the CFPs derived from first principles, the DMFT results of Ref. \cite{pourovskii} provide the closest fit at 5 K, which can be attributed to the large $A_{6}^6\langle r^6 \rangle$ parameter obtained in that work. However, the exchange field derived in Ref. \cite{pourovskii} is not sufficiently large to reproduce the positions of the peaks in the spectrum, even though it was shown to produce magnetization properties consistent with experiment.

\begin{figure}[t]
    \centering
    \includegraphics[width=0.45\textwidth]{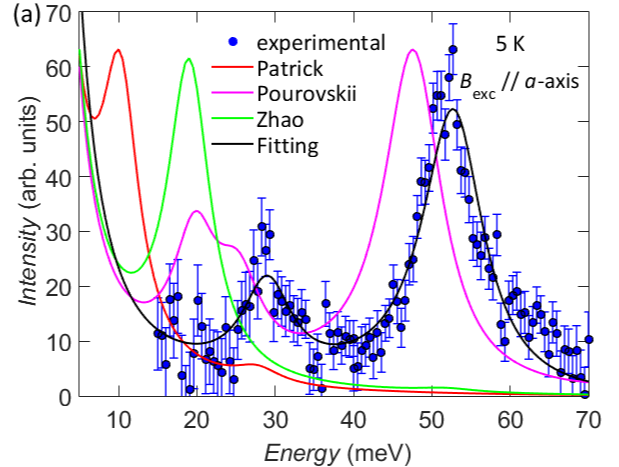}
    \includegraphics[width=0.45\textwidth]{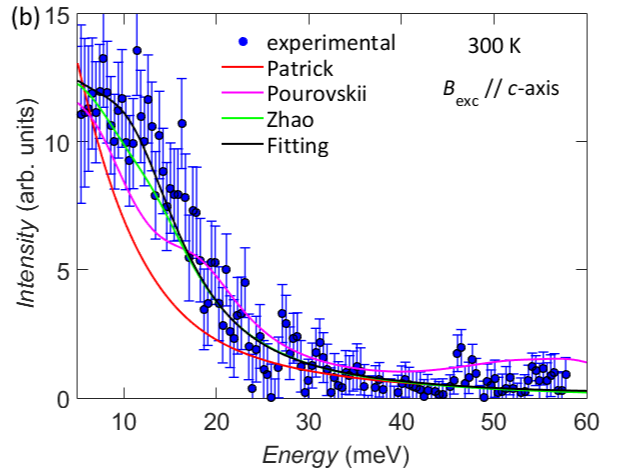}
    \caption{Fitting to the experimental spectrum using \texttt{PyCrystalField} and calculated neutron spectra using different parameter sets from previous references \cite{patrick, zhao, pourovskii} compared with the experimental data at (a) 5~K and (b) 300~K, both measured with a neutron incident energy of 80 meV.}
    \label{fig:comp_spectrum}
\end{figure}

\begin{table*}[ht]
\caption{\label{tab:CFPs}
NdCo$_5$ crystal-field parameters and magnitude of the exchange field, whose direction is along the $a$-axis, obtained by fitting to the CEF transitions observed experimentally at 5 K. The CFPs are in the Stevens notation. We compare the experimental values obtained here with some previous theoretical works. The parameters obtained by Zhao \textit{et al} were obtained by fitting magnetization curves.
}
\begin{ruledtabular}
\begin{tabular}{lccccc}
\textrm{}&
\textrm{A$_2^0\langle r^2\rangle$ (K)}&
\textrm{A$_4^0\langle r^4\rangle$ (K)}&
\textrm{A$_6^0\langle r^6\rangle$ (K)}&
\textrm{A$_6^6\langle r^6\rangle$ (K)}&
\textrm{B$_{\textrm{exc}}$ (T)}\\
\colrule
custom \texttt{python} code & -300 & 0 & 10 & 900 & 535\\
\texttt{SPECTRE} & -537$\pm$21 & 0 & 0& 913$\pm$55 & 535$\pm$2\\
\texttt{PyCrystalField} & -243 & 0 & 0 & 1160 & 470\\
Patrick and Staunton \cite{patrick} & -415 & -26 &5 & 146 & 252\\
Pourovskii \textit{et al}. \cite{pourovskii} & -285 & -33 & 36 &1134 & 292\\
Zhao \textit{et al}. \cite{zhao} & -510 & 0 & 7& 143 & 558\\
\end{tabular}
\end{ruledtabular}
\end{table*}

Having obtained a set of parameters that reproduces the experimentally observed CEF excitations at $5$~K, we now check the agreement of these parameters with our results at high temperature. At $300$~K, above the SRT, the NdCo$_5$ magnetization easy-axis is parallel to the crystallographic $c$-axis. Rotating the axis of the exchange field and assuming that the CFPs and the $|B_{\textrm{exc}}|$ do not change significantly at 300 K, we obtain the spectrum shown in Fig. \ref{fig:comp_spectrum}b considering a FWHM of 10 meV for the peaks at this temperature. We can see that the parameter set obtained by fitting the 5 K data also agrees well with the experimental spectrum at 300 K, showing no peaks above 25 meV. However, due to thermal broadening of the peaks and the experimental resolution close to the elastic line, it was not possible to experimentally resolve the peaks below 25 meV.

\subsection{Comparison of CFPs}
All previous studies agree that the A$_2^0\langle r^2\rangle$ is negative, which drives the basal plane anisotropy. The A$_2^0\langle r^2\rangle$ value obtained by SPECTRE is somewhat larger in magnitude than the values calculated from first principles in  refs.~\cite{patrick,pourovskii} but consistent with the larger value obtained in Ref.~\cite{zhao} based on the fitting of magnetization data. We further find that both A$_4^0\langle r^4\rangle$ and A$_6^0\langle r^6\rangle$ do not significantly affect
the excitation spectrum, so they cannot be strongly constrained by our measurements. The A$_4^0\langle r^4\rangle$ in the custom \texttt{python} code was introduced to improve the agreement with the magnetization data. However, the spectrum is sensitive to both A$_6^6\langle r^6\rangle$ and $B_{\textrm{exc}}$, leading to the best fit values of 1160~K and 470~T respectively.
Pourovskii et al.~\cite{pourovskii} found a similarly large value of $A_6^6\langle r^6\rangle$ using their DMFT framework, and argued that this term also accounts for the nonsaturated Nd moments at zero temperature. Relatively large values of $l=6$ coefficients are also found necessary to explain, e.g.\ the spin-reorientation in Nd$_2$Fe$_{14}$B \cite{herbst}.
Calculations which do not explicitly include hybridization of the $4f$ electrons with their environment (such as the yttrium-analogue model of Ref.~\cite{patrick}) do not produce these large higher-order CFPs. The analysis of our inelastic neutron spectra thus corroborates the idea that a standard assumption of crystal field theory --- that the strongly localized $f$-electrons do not themselves affect the crystal field --- does not hold in NdCo$_5$.

We finally note that the $B_{\textrm{exc}}$ extracted from the fit is considerably larger than the estimations from Refs. \cite{patrick,pourovskii}. This, together with the large $A_2^0\langle r^2 \rangle$ is expected to cause an overestimation of the SRT temperatures $T_{SR1}$ and $T_{SR2}$ compared to both calculations of the SRT using previous parameter sets and experiments. The former, however, do not consider a range of possible additional terms in the Hamiltonian, including exchange anisotropy and the anisotropy on the Co sites \cite{Zhu2014}. Some of these terms can compensate the influence of $B_{\textrm{exc}}$ and $A_2^0\langle r^2 \rangle$ and restore the predicted $T_{SR1}$ and $T_{SR2}$ to the experimentally observed temperatures.  

\section{Conclusion}
We have performed inelastic neutron scattering to investigate the crystal electric field (CEF) levels in the intermetallic ferrimagnets RECo$_{5}$ (RE = Nd and Y). The large $A_6^6\langle r^6\rangle$ extracted from our experimental data as well as the large line-widths of the inelastic peaks highlight the importance of the interaction between the localized $f$-electrons and itinerant $d$-electrons for both the CEF and magnetic anisotropic interactions. The former is in good agreement with previous calculations \cite{pourovskii}, although the exchange field is considerably higher than previously reported values. In light of the ongoing discussion around the magnetism of other technologically relevant rare-earth intermetallics, including the Nd$_2$Fe$_{14}$B family \cite{Bouaziz2023,Boust2022}, we are hopeful that our approach to fitting full inelastic neutron CEF spectra can help to shed further light on the interplay of interactions that generate their interesting magnetic properties.

\begin{acknowledgments}
The authors acknowledge A. Scheie for helping with the implementation of the exchange field interaction in \textrm{PyCrystalField}. J.L.J acknowledges FAPESP-Young Investigator Grant 2018/08845-3 and CNPq 31005/2021-6. J.L.J, G.J.N and F.P acknowledge FAPESP 2019/24797-1 and 2019/24711-0. G.J.N also acknowledges Print CAPES 2022/33002010002P2. The work at the university of Warwick was supported through Grants EP/M02941/1 and EP/T005963/1 from EPSRC, UK.

\end{acknowledgments}

\section{Appendixes}
Table \ref{tab:eigenvectors} show the eigenvalues with respective eigenvectors in the $|SLJm_J\rangle$ basis obtained by \texttt{PyCrystalField} with the quantization axis along the $a$ axis. The eigenvectors in the J-basis were obtained re-scaling the CFPs by $\Theta_k^J/\Theta_k^{LS}$, where $\Theta_k^J$ and $\Theta_k^{LS}$ are the Stevens factors in the J-basis and LS-basis, respectively, both already implemented in \texttt{PyCrystalField}. The exchange field term in the J-basis Hamiltonian was also modified to couple with the J operator using the following expression: $\mathcal{H}_{\textrm{exc}}=2(g_J-1)\mu_B\mathbf{B}_{\textrm{exc}}\cdot\mathbf{J}$.

The CFPs in Table \ref{tab:CFPs} are in the Stevens notation as originally derived by Stevens \cite{stevens}. Although \texttt{PyCrystalField} also uses the Stevens convention, it defines the CFPs as $B_k^q=A_k^q\langle r^k\rangle \Theta_k$, where $\Theta_k$ are the Stevens factors \cite{stevens}. On the other hand, \texttt{SPECTRE} uses the CFPs $W_k^q=A_k^q\langle r^k\rangle/\lambda_{kq}$ in the Wybourne notation, where $\lambda_{kq}$ are tabulated factors \cite{spectre}. Tables \ref{tab:pycrystalfield} and \ref{tab:spectre} show the CFPs as obtained by \texttt{PyCrystalField} and \texttt{SPECTRE}, respectively, prior to performing any conversion.

\begin{table*}
\caption{Eigenvalues and eigenvectors in the $|J=9/2\rangle |m_J \rangle$ basis for the parameters set in Table \ref{tab:CFPs} obtained using the \texttt{PyCrystalField} software. Note that here the quantization axis was chosen along the $a$ axis and the exchange field along the $c$ axis, as adopted in other references. To obtain these eigenvectors, we rotated the CFPs in table \ref{tab:CFPs} from $z||c$ to $z||a$ using the rotation matrices in Ref. \cite{rudowicz}.} 

\begin{ruledtabular}
\begin{tabular}{c|cccccccccc}
E (meV) &$| -\frac{9}{2}\rangle$ & $| -\frac{7}{2}\rangle$ & $| -\frac{5}{2}\rangle$ & $| -\frac{3}{2}\rangle$ & $| -\frac{1}{2}\rangle$ & $| \frac{1}{2}\rangle$ & $| \frac{3}{2}\rangle$ & $| \frac{5}{2}\rangle$ & $| \frac{7}{2}\rangle$ & $| \frac{9}{2}\rangle$ \tabularnewline
 \hline 
0.000 & 0.0 & -0.0017 & 0.0 & 0.0037 & 0.0 & -0.0338 & 0.0 & -0.2731 & 0.0 & -0.9614 \tabularnewline
27.250 & 0.0 & -0.0101 & 0.0 & 0.0619 & 0.0 & 0.1599 & 0.0 & -0.9491 & 0.0 & 0.2642 \tabularnewline
28.165 & 0.0009 & 0.0 & -0.0331 & 0.0 & 0.031 & 0.0 & 0.5365 & 0.0 & -0.8427 & 0.0 \tabularnewline
53.443 & 0.005 & 0.0 & -0.0663 & 0.0 & 0.1252 & 0.0 & 0.8317 & 0.0 & 0.5368 & 0.0 \tabularnewline
78.671 & 0.0 & -0.0522 & 0.0 & 0.3558 & 0.0 & 0.9167 & 0.0 & 0.1572 & 0.0 & -0.0754 \tabularnewline
90.180 & 0.0027 & 0.0 & 0.526 & 0.0 & -0.8375 & 0.0 & 0.1426 & 0.0 & 0.0393 & 0.0 \tabularnewline
90.509 & 0.0 & 0.2224 & 0.0 & -0.9046 & 0.0 & 0.3632 & 0.0 & -0.0044 & 0.0 & -0.0154 \tabularnewline
102.573 & 0.2641 & 0.0 & 0.8169 & 0.0 & 0.5126 & 0.0 & -0.0038 & 0.0 & -0.0153 & 0.0 \tabularnewline
138.368 & -0.9645 & 0.0 & 0.2248 & 0.0 & 0.1387 & 0.0 & 0.0042 & 0.0 & -0.0021 & 0.0 \tabularnewline
142.132 & 0.0 & -0.9735 & 0.0 & -0.2264 & 0.0 & 0.0323 & 0.0 & 0.0009 & 0.0 & -0.0005 \tabularnewline
\end{tabular}\end{ruledtabular}
\label{tab:eigenvectors}
\end{table*}

\begin{table}[h]
\caption{\label{tab:pycrystalfield}
NdCo$_5$ crystal-field parameters in the Stevens convention and magnitude of the exchange field obtained by fitting the experimental spectrum at 5 K using \texttt{PyCrystalField} in the intermediate-coupling scheme.
}
\begin{ruledtabular}
\begin{tabular}{ccc}
\textrm{B$_2^0$ (meV)}&
\textrm{B$_6^6$ (meV)}&
\textrm{B$_{\textrm{exc}}$ (T)}\\
\colrule
0.08458726 & -0.00112046 & 470\\
\end{tabular}
\end{ruledtabular}
\end{table}

\begin{table}[h]
\caption{\label{tab:spectre}
NdCo$_5$ crystal-field parameters in the Wybourne convention and magnitude of the exchange field obtained by fitting the experimental spectrum at 5 K using \texttt{SPECTRE} software in the intermediate-coupling scheme.
}
\begin{ruledtabular}
\begin{tabular}{ccc}
\textrm{W$_2^0$ (meV)}&
\textrm{W$_6^6$ (meV)}&
\textrm{B$_{\textrm{exc}}$ (T)}\\
\colrule
-93 & 83 & 535\\
\end{tabular}
\end{ruledtabular}
\end{table}

\bibliography{main}

\end{document}